\documentclass[graybox]{svmult}

\usepackage{type1cm}        
\usepackage{makeidx}         
\usepackage{graphicx}        
\usepackage{mathrsfs}   
\usepackage{multicol}        
\usepackage[bottom]{footmisc}
\usepackage{color}

\usepackage{amsmath}
\usepackage{bm}
\usepackage{txfonts}
\renewcommand\mathbf{\bm}

\makeindex             

\begin{document}

\title*{Strongly Interacting Matter Under Rotation: An Introduction}
\author{Francesco Becattini, Jinfeng Liao, and Michael Lisa}
\institute{Francesco Becattini \at University of Florence \\ \email{becattini@fi.infn.it } 
\and Jinfeng Liao \at Physics Department and Center for Exploration of Energy and Matter, Indiana University, 2401 N Milo B. Sampson Lane, Bloomington, IN 47408, USA,\\ \email{liaoji@indiana.edu} 
\and Michael Lisa \at  Department of Physics, The Ohio State University,
191 West Woodruff Avenue, Columbus, OH 43210 USA.\\ \email{lisa.1@osu.edu}}
%
%
\maketitle

\abstract{ Ultrarelativistic collisions between heavy nuclei briefly generate the quark-gluon
plasma (QGP), a new state of matter characterized by deconfined partons last seen microseconds
after the Big Bang.
The properties of the QGP are of intense interest, and a large community has developed over
several decades, to produce, measure and understand  this primordial plasma.
The plasma is now recognized to be a strongly-coupled fluid with remarkable properties,
and hydrodynamics is commonly used to quantify and model the system.
An important feature of any fluid is its vorticity, related to the local
angular momentum density; however, this degree of freedom has received relatively little
attention because no experimental signals of vorticity had been detected.
Thanks to recent high-statistics datasets from experiments with precision tracking and complete
kinemetic coverage at collider energies, hyperon spin polarization measurements have begun
to uncover the vorticity of the QGP created at the Relativistic Heavy Ion Collider.
The injection of this new degree of freedom into a relatively mature field of research
represents an enormous opportunity to generate new insights into the physics of the QGP.
The community has responded with enthusiasm, and this book 
 (to be published as a volume of Lecture Notes in Physics series by Springer)
 represents some of the
diverse lines of inquiry into  aspects of strongly interacting matter under rotation.
}

\newpage
\section{Milestones}\label{sec:Milestones}


In 2005, Liang and Wang~\cite{Liang:2004ph} predicted that spin-orbit coupling would
  polarize strange quarks created in non-central heavy ion collisions, resulting in
  emitted $\Lambda$ hyperons globally polarized along the direction of the collision angular momentum.
The magnitude and momentum-dependence of the predicted polarization depended on details
  of specific models of quark-quark potentials, small-angle scattering approximations,
  and details of hadronization mechanisms.
  
In 2008, Becattini and collaborators~\cite{Becattini:2007sr} noted that in a hydrodynamic
picture, local thermodynamic equilibrium implies a relation between the spin polarization
and the rotational flow structure (vorticity). In the hydrodynamic model, vorticity can be 
extracted directly from the evolution, with no need to appeal to specific microscopic processes.
In 2013, an equation relating  the polarization of $\Lambda$ hyperons and 
thermal vorticity was derived \cite{Becattini:2013fla} and such polarization was predicted to be at the level of 
a few percent. 
The first result  regarding the systematic dependence of this effect on the collision beam energy, particularly in the range relevant to the beam energy scan program at the Relativistic Heavy Ion Collider (RHIC), was reported in a 2016 paper~\cite{Jiang:2016woz}, providing a highly relevant insight for the later experimental measurements.

In 2017, the STAR Collaboration published~\cite{STAR:2017ckg} the first observation of
 global $\Lambda$ polarization from noncentral heavy ion collisions.
As discussed below and throughout this Volume, most theoretical interpretations
  of these observations are based upon this hydrodynamic approach.

While the phenomenon of global polarization was predicted 
  based on particle-particle interaction, the success of quantitative predictions of the 
  hydrodynamic model to reproduce experimental observations (discussed below) seem to
  confirm that for spin, as for many other observables, 
  microscopic details are less important than bulk thermodynamic properties.
Below, we discuss the hydrodynamic approach to vorticity and polarization, followed
  by experimental observations.
  
We will briefly discuss the related phenomenon of vector meson spin alignment, also
  predicted by Liang and Wang~\cite{Liang:2004xn} in 2005.
As of now, measurements of spin alignment 
  at the Large Hadron Collider (LHC) and RHIC are difficult to explain
  in any theoretical approach.

\section {Introduction}\label{sec:intro}

Twenty years ago, the world's first nuclear collider began producing heavy ion collisions at
  energies far surpassing those previously achievable in fixed-target experiments.
The goal was to produce the quark-gluon plasma (QGP)-- a state of matter characterized
  by partonic (rather than hadronic) degrees of freedom.
For decades, production and study of the QGP had long been the focus driving the field
  of relativistic heavy ion physics, as it holds the promise of shedding light on the
  non-perturbative region of quantum chromodynamics (QCD), the most poorly understood of 
  the fundamental interactions in the Standard Model. 
   
In 2005~\cite{Adams:2005dq,Adcox:2004mh,Back:2004je,Arsene:2004fa}, 
  based on a systematic and comprehensive analysis of available data,
  the experimental collaborations at the Relativistic Heavy Ion
  Collider (RHIC) confirmed that QGP is indeed created in ultra-high energy collisions.
Furthermore, the data clearly indicated that the QGP was a strongly coupled fluid, contrary
  to some expectations that the plasma would be weakly coupled due to the combination of
  high temperatures and the running of the QCD coupling constant.
The evidence driving this conclusion was the collective anisotropic emission distribution
  of hadrons from the collision-- the so-called ``elliptic flow.''
These very strong anisotropies (and the dependence upon mass and momentum) were nearly
  quantitatively consistent with expectations based on relativistic inviscid (ideal) 
  hydrodynamics.

The discovery of nearly ``perfect fluid'' behavior had two major outcomes.
Firstly, it prompted a re-evaluation of numerical QCD calculations performed
  on a lattice, the most reliable {\it ab initio} calculations of the
  strong interaction.
While numerically correct, lattice calculations could be misinterpreted to suggest
  that a weakly coupled gas of quarks and gluons was the proper
  paradigm for modeling collisions at RHIC. 
It was also realized that the QGP near the pseudo-critical transition
temperature is a peculiar system: unlike ordinary matter, its microscopic 
interaction length is comparable to the  thermal de Broglie wavelength, making the kinetic collisional 
description inappropriate. Nevertheless, even under such unusual conditions, the local thermodynamic
equilibrium concept and hydrodynamics are still valid,.  Hence, the discovery established 
relativistic fluid dynamics as the new paradigm for the bulk evolution of the system.
Confronting increasingly sophisticated hydrodynamic calculations with
  data has produced valuable estimates of transport coefficients,
  initial parton distributions, and the QCD equation of state.
Triangular and higher-order azimuthal correlations have probed
  the substructure of the fluid flow fields at ever finer scale.

A relativistic collision between heavy nuclei at finite impact parameter can 
involve  angular momentum of order $10^{3\sim 5}\hbar$. In a fluid, angular momentum can 
manifest as vorticity, rotational gradients of the flow and temperature fields~\cite{Becattini:2007sr}.
Until recently, this aspect of the plasma had been largely
ignored, as there had been no experimental observation of its effects.
  
In 2017, the STAR Collaboration published an observation of 
global hyperon polarization in Au+Au collisions at 
RHIC, opening the potential to probe novel substructures of the QGP fluid at the 
finest possible scale. This is a rare case in which an entirely new direction is introduced
to a mature field. It is especially exciting because the natural language for discussing
vorticity-- three-dimensional relativistic viscous hydrodynamics-- has been developed to 
a high degree of sophistication by a large community of theorists.
It is an opportunity for new insights into the physics of deconfined
  QCD matter, and the heavy ion community has responded with intense
  focus on the topic.
This book represents a broad sampling of directions of inquiry into
  this new area of research.

\section{Accessing subatomic vorticity}
\label{sec:data}

``Lumpy" azimuthal fluid flow patterns (elliptic flow, triangular flow, etc) may be measured by azimuthal correlations
  between the momenta of emitted particles; this is experimentally straight-forward.
Orbital angular momentum in heavy ion collisions, on the other hand, is experimentally inaccessible.
Instead, one relies on coupling between the orbital (``mechanical'') angular momentum of the fluid 
  and spin of the emitted particles.
The first observation of such an effect was reported more than a century ago by S. Barnett~\cite{PhysRev.6.239},
  in which an uncharged and un-magnetized solid metal object, when set spinning, spontaneously
  magnetizes\footnote{That the magnetization arose from spin polarization of the electrons was
  not known to Barnett and his contemporaries in 1915, as the concept of quantum spin 
  was not introduced until nearly a decade later. }.
 
The analogous effect in a {\it fluid}, coupling mechanical vorticity of the bulk
  fluid and quantum spin polarization, was first reported by Takahashi et al, in 2016~\cite{TakahashiFluidSpintronics}.
In their experiment, liquid mercury flowing through a channel acquired local vorticity
  due to viscous friction with the wall.
Spin-vorticity coupling produced a polarization gradient that could then be detected
  directly through the inverse spin Hall effect.
 The results could be understood by expanding angular momentum conservation
  in fluid dynamics, to include angular momentum transfer between the liquid and electron 
  spin~\cite{TakahashiFluidSpintronics}.

In the Barnett and Takahashi experiments, the macroscopic rotational motion was a
  controlled variable and the spin polarization straightfoward to measure.
In high-energy nuclear collisions, the magnitude and direction of the angular momentum
  fluctuates from one event to the next, and a statistically significant measurement
  requires combining $\sim10^7-10^8$ events.
Furthermore, the particles whose polarization is to be measured are emitted at 
  all angles at speeds approaching that of light.

\begin{figure}[t]
    \centering
    \includegraphics[width=0.95\textwidth]{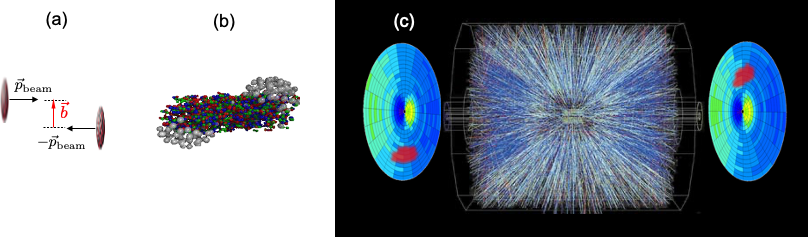}
    \caption{The geometry of a collision.  (a) Before collision: the angular momentum is determined by the impact
    parameter, $\vec{b}$, an uncontrolled variable that fluctuates from one collision to the next.
    (b) In a non-central ($|\vec{b}|\neq0$) collision, parts of the nuclei overlap, producing the QGP, 
    while the so-called ``spectators'' continue to travel forward, experiencing only a slight impulse
    directed away from the collision.  (c) One reconstructed event in two subsystems in STAR experiment.
    The Time Projection Chamber (TPC)~\cite{Anderson:2003ur} 
    records $\sim10^3$ charged particles emitted from the QGP created in
    the collision, while the Event Plane Detector (EPD)~\cite{Adams:2019fpo} measures spectator fragments.
    The magnitude and direction of $\vec{b}$ are determined, respectively, by the number of charged particles measured
    in the TPC and the anisotropic hit pattern in the EPD.}
    \label{fig:geometry}
\end{figure}

These challenges are addressed by precision tracking and correlating detector
  subsystems in different regions of the experiment.
In particular, the angular momentum in a collision is given by
\begin{equation}
    \vec{J} = \vec{b}\times\vec{p}_{\rm beam} ,
\end{equation}
where the impact parameter, $\vec{b}$, is the transverse (to the beam direction) 
  vector connecting the center of the target nucleus to that of the beam
  nucleus (where attention to the designation of beam and target is important~\cite{Abelev:2007zkERRATUM}),
  and $\vec{p}_{\rm beam}$ is the momentum of the beam in the collision center-of-momentum (c.o.m.) frame.
The magnitude of the impact parameter, $|\vec{b}|$, is estimated by the total number of charged particles emitted roughly
  perpendicular to the beam in the collision c.o.m. frame,  while its direction, $\hat{b}$, is estimated
  by the sidewards deflection of particles emitted close to the beam direction.
See figure~\ref{fig:geometry} for an illustration.

The flow pattern of the QGP fluid is complex and any local vorticity may fluctuate
  as a function of position within each droplet; however, the {\it average} vorticity 
  must be parallel to $\vec{J}$ which is event-specific.
For this reason, spin polarization projection along $\hat{J}$ is termed the
  ``global'' polarization.

Having determined the direction\footnote{In principle, the magnitude $|\vec{J}|$ of the collision's
angular momentum may be estimated as well.  However, not all of this angular momentum is transferred
to the plasma at midrapidity~\cite{Jiang:2016woz}, so usually only the direction $\hat{J}$ is of interest.
This quantity is the only important ingredient to estimate vorticity in any event.} of the average
vorticity, the second challenge is to measure the spin polarization along that direction.

\begin{figure}[t]
    \centering
    \includegraphics[width=0.8\textwidth]{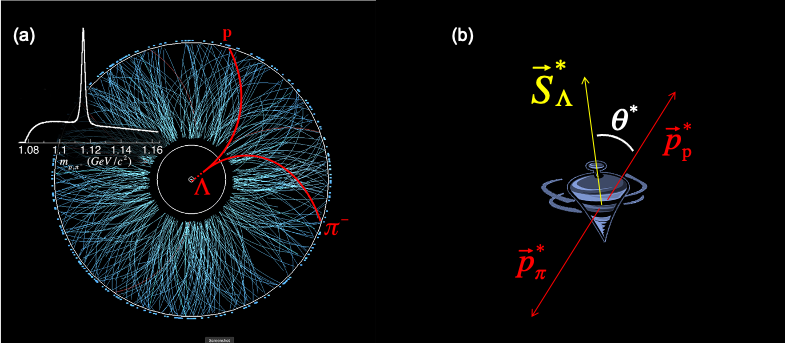}
    \caption{(a) A $\Lambda$ hyperon detected in the STAR TPC by combining its charged proton and pion daughters.
    Inset: the invariant mass of daughter pairs shows a clear peak at the $\Lambda$ mass.  (b) In the parity-violating
    decay topology, the daughter proton tends to be emitted in the direction of the parent $\Lambda$ hyperon, in
    the $\Lambda$ center of mass frame.}
    \label{fig:AlexLambda}
\end{figure}

If the QGP fluid does indeed have non-vanishing vorticity, and if thermalization (complete or
  partial) of orbital and spin degrees of freedom does occur, then presumably all particles
  emitted in the collision will have their average spins aligned with $\hat{J}$.
Of the zoo of particle types emitted in a heavy ion collision, the spin directions of only a few
  are easily measurable.
In particular, particles undergoing parity-violating weak decay betray their spin direction
  through asymmetries in the momentum distribution of their daughters.
Of this already restricted subset of particles, only a few are created in reasonable numbers
  to allow a significant measurement.
The best candidate is the $\Lambda$ hyperon, which can be cleanly measured by its $p+\pi^-$
  decay in the TPC, as seen in panel (a) of figure~\ref{fig:AlexLambda}.
The decay topology is sketched in panel (b) of figure~\ref{fig:AlexLambda}.
An ensemble with polarization $\vec{P}_{\Lambda}$ will preferentially emit
   daughter protons along the direction of polarization according to
\begin{equation}
\label{eq:SelfAnalyzing}
\frac{dN}{d\cos\theta^*}
= \tfrac{1}{2}\left(1+\alpha_{\Lambda}\vec{P}_{\Lambda}\cdot\hat{p}_{p}^*\right) ,
\end{equation}
where $\theta^*$ is the angle between the polarization and daughter proton momentum $\vec{p}^*_{p}$
in the hyperon rest frame.  The decay parameter
$\alpha_{\Lambda}=0.732$ determines the strength of the effect.

The global polarization is then measured by correlating information from both detector
  subsystems:
  \begin{equation}
  \label{eq:cumulant}
      \left\langle \vec{P}_{\Lambda}\cdot\hat{J}\right\rangle 
              =\frac{8}{\pi\alpha_H R_{\rm EP}^{(1)}}\left\langle\sin\left(\Psi_{\rm EP,1}-\phi^*_p\right)\right\rangle ,
  \end{equation}
where $\phi^*_p$ is the azimuthal angle of the daughter proton in the parent hyperon frame.
In equation~\ref{eq:cumulant},
  $\Psi_{EP,1}$ is the first-order event plane angle, an estimator of the azimuthal angle of the 
  impact parameter $\vec{b}$; the resolution of this estimation is $R_{EP}^{(1)}$.
Standard methods have been developed to extract both the event plane and the resolution from anisotropic
  particle distributions in the EPD.

\begin{figure}[t]
    \centering
    \includegraphics[width=0.8\textwidth]{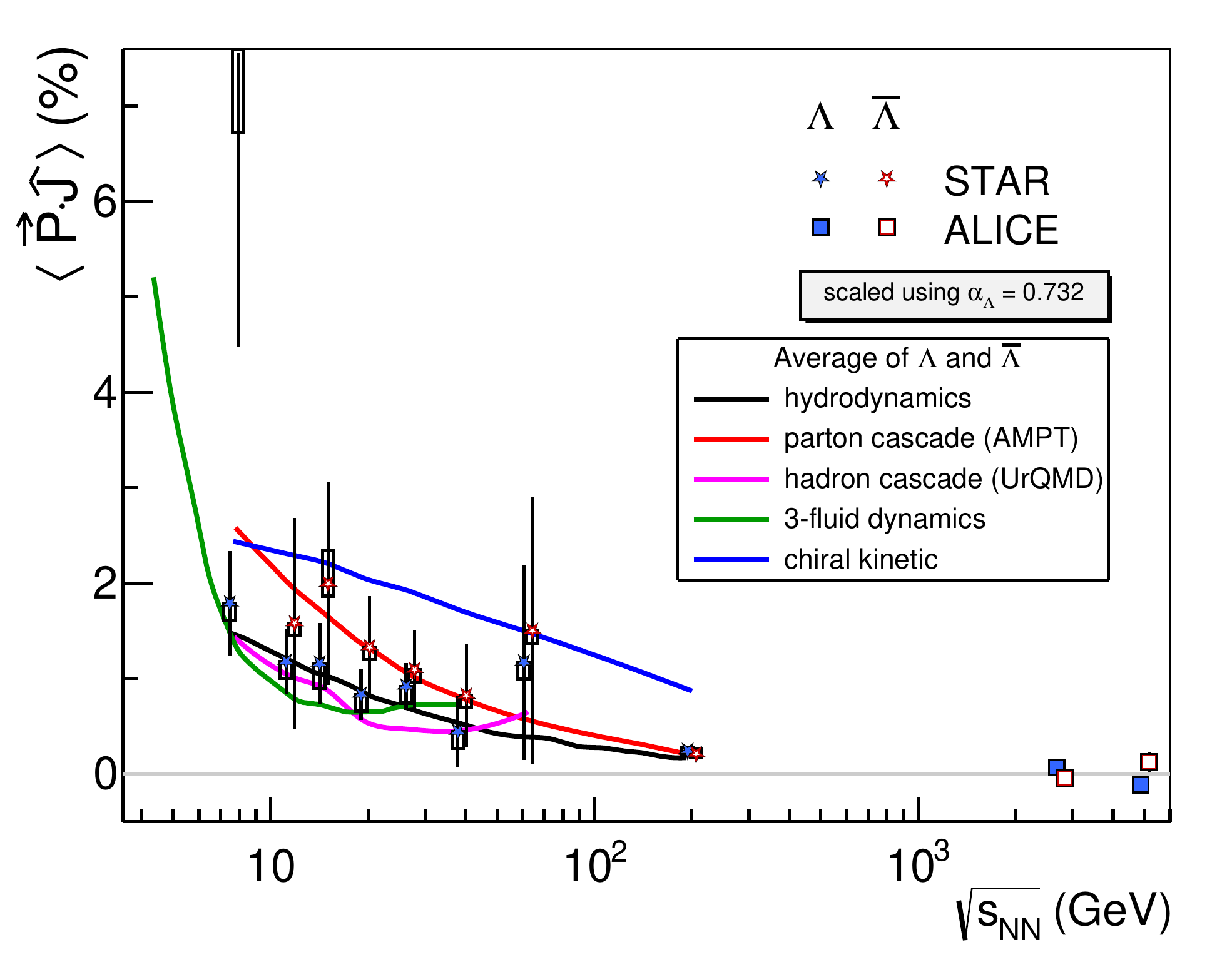}
    \caption{The world dataset of global $\Lambda$ hyperon polarization in relativistic heavy
    ion collisions compared to expectations from hydrodynamic and transport simulations.
    Figure from~\cite{Becattini:2020ngo}.}
    \label{fig:GlobalVSRoots}
\end{figure}

The discussion thus far has described the global $\Lambda$ hyperon polarization measurement in the STAR
  experiment at RHIC.
The ALICE experiment at the LHC performed a similar analysis, tracking charged hyperon daughters
  with a gas-filled TPC at midrapidity, and measuring $\Psi_{\rm EP,1}$
  with segmented detectors at forward rapidity.
The STAR and ALICE measurements thus far comprise the world's dataset on global polarization,
  and are shown in figure~\ref{fig:GlobalVSRoots}.

We offer some general remarks on figure~\ref{fig:GlobalVSRoots} in the next section,
  but at the experimental level, we note that the statistical uncertainties at low $\sqrt{s_{NN}}$
  are large.
These uncertainties are determined by (1) the number of collision events recorded by the experiment;
  (2) the per-event hyperon yield; (3) the event-plane resolution $R_{\rm EP}^{(1)}$.
Measurements by the STAR Collaboration in the second phase of the RHIC Beam Energy Scan (BES-II)~\cite{Bzdak:2019pkr}
  will have an order of magnitude better statistics~\cite{Aggarwal:2010cw} and better event plane resolution~\cite{Adams:2019fpo}; overall, the precision should increase roughly eight-fold,
  allowing important systematic studies~\cite{Becattini:2020ngo} not currently possible.

\begin{figure}[t]
    \centering
    \includegraphics[width=0.8\textwidth]{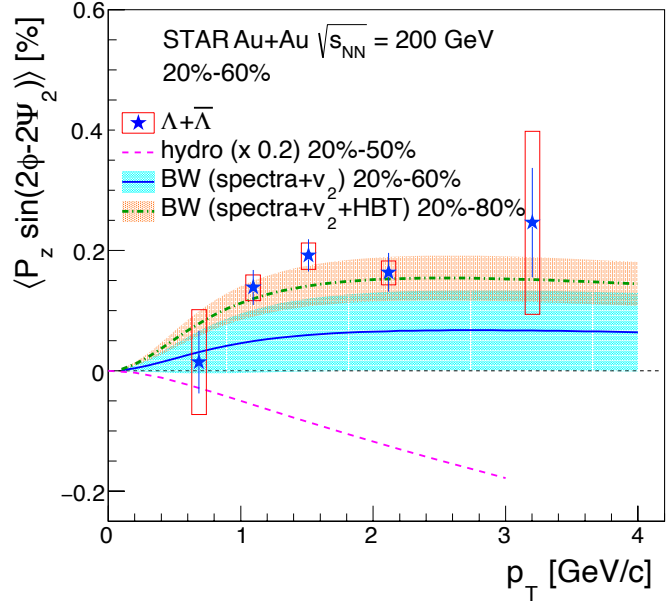}
    \caption{The second-order Fourier coefficient of the azimuthal oscillation of the longitudinal component of the
      $\Lambda$ hyperon polarization.  Figure from~\cite{Adam:2019srw}.}
    \label{fig:PzSTAR}
\end{figure}

The average ``global''   polarization vector must point along the direction of $\hat{J}$. On the
other hand, the mean spin polarization vector for particles with specific momentum have 
three components which can be also measured. The component along the beam (longitudinal component)
is expected to show a $2^{\rm nd}$-order azimiuthal oscillation relative to the event plane.
The amplitude and phase of this oscillation has been measured for Au+Au collisions at $\sqrt{s_{NN}}=200$~GeV
  by the STAR Collaboration.
Figure~\ref{fig:PzSTAR} shows the transverse momentum dependence of the $2^{\rm nd}$ Fourier component
  for non-central collisions.
Thanks to the excellent tracking, good event plane resolution, and a high statistics dataset available at 
  RHIC top energy, an oscillating sub-percent
  polarization signal is easily measured.

\section{From signal to physics}
\subsection{Hydrodynamics as the basis to understand hyperon polarization}

As we discussed in section~\ref{sec:Milestones}, the original idea for global $\Lambda$
 polarization in heavy ion collisions was based on microscopic processes that drove
  the initial state, the transfer of angular momentum from orbital to spin degrees of
  freedom, and the subsequent hadronization mechanism.
Assumptions and parameters were required to compute each of these components of the
  calculation. 
  A detailed discussion along this line can be found in Chapter 7~\cite{Gao:2020lxh}.
  
The tremendous success of hydrodynamics to heavy ion physics suggests that  
the  myriad details of microscopic processes undoubtedly at play in these complex collisions
  are eventually unimportant, as the system approaches local equilibrium quickly.
In the earliest days of RHIC, ideal (inviscid), boost-invariant hydrodynamic calculations
  with simple initial conditions largely reproduced-- nearly ``out of the box"-- the multiplicity,
  $p_T$ and mass systematics of measured elliptic flow.
This success gave some confidence that equilibrium hydrodynamics was a good paradigm
  to understanding the collective physics of heavy ion collisions.

In the subsequent decades, several important insights have been achieved by working within
  this framework, using {\it details} in the data to probe the partonic structure of the
  initial state, transport coefficients, and hadronization mechanisms.
These insights required considerable elaboration of the initial simple models, incorporating
  viscosity, baryochemical currents, vorticity, three-dimensional dynamics, and event-by-event
  fluctuations in the initial state. 
However,  the close resemblance of the initial simple calculations with observations set
  this fruitful enterprise on firm ground.
  
Figure~\ref{fig:GlobalVSRoots} suggests that the same situation exists in the study of
global hyperon polarization. Theoretical curves show predictions from hydrodynamic and transport calculations, in which fluid vorticity is assumed to equilibrate with $\Lambda$
spin degrees of freedom to produce the polarization.
Vorticity -- more properly, thermal vorticity-- is 
calculated directly from the flow field in the hydrodynamic calculations,
as discussed in detail in Chapter 8~\cite{Karpenko:2021wdm}. 
On the other 
hand, in the transport calculations, flow and temperature fields are calculated from the motion
of multiple particles in coarse-grained spatial cells; this implicitly assumed local
thermalization; see detailed discussions in Chapter 9~\cite{Huang:2020dtn}. 
Eventually, in both methods, polarization of a spin 1/2
fermion is obtained from the same formula relating mean spin to the thermal vorticity 
$\varpi$ at the leading order \cite{Becattini:2013fla}:
\begin{equation}\label{spinvector}
 S^\mu(p) = -\frac{1}{8m} \epsilon^{\mu\nu\rho\sigma} p_\sigma 
 \frac{\int_\Sigma {\rm d}\Sigma \cdot p \varpi_{\nu\rho} n_F (1-n_F)}
 {\int_\Sigma {\rm d}\Sigma \cdot p n-F}
\end{equation} 
where $S^\mu(p)$ is the mean spin vector and $n_F$ is the covariant Fermi-Dirac distribution 
function. The thermal vorticity is defined as the antisymmetric derivative of the four-temperature
vector field, that is:
\begin{equation}
 \varpi_{\mu\nu} = \frac{1}{2} \left[ \partial_\nu \left(\frac{1}{T} u_\mu \right) -
 \partial_\mu \left(\frac{1}{T} u_\nu \right) \right]
\end{equation}
where $T$ and $u_\mu$ are local temperature field and flow velocity field, respectively.
The integration in equation~\eqref{spinvector} 
is performed 
on the 3-D hadronization hypersurface $\Sigma$. The 
polarization vector $P^\mu$ is simply $S^\mu/|S|$ and its global, momentum integrated, value
in the particle rest frame turns out to be directed along the angular momentum vector, so
that, approximately one has:
\begin{equation}
    \left\langle\vec{P} \right\rangle \cdot\hat{J} \approx \frac{1}{2}\left\langle\varpi
    \right\rangle \cdot \hat{J}
\end{equation}
where the $\left\langle\varpi \right\rangle$ is the mean thermal vorticity value over the
hadronization hypersurface.  
The above relation is a direct manifestation for the rotational polarization of microscopic spin. The theoretical underpinning of this phenomenon is to be fully elaborated  through a variety of approaches such as quantum field theory (in Chapters 2~\cite{Becattini:2020sww}, 3~\cite{Buzzegoli:2020ycf} and 4~\cite{Ambrus:2019cvr}) and relativistic kinetic theory (in Chapters 5~\cite{Tinti:2020gyh} and 6~\cite{ch:Jiang}).

For the most part, these models have been used to understand other observations from heavy
 ion collisions, and the results in figure~\ref{fig:GlobalVSRoots} are obtained largely
  ``out of the box". 
The quantitative agreement, as well as the universal decreasing trend of polarization with
 $\sqrt{s_{\rm NN}}$ (despite the fact that $|\vec{J}|$ {\it in}creases with
  increasing collision energy) is a clear indication that we have at hand a paradigm
  to understand hyperon polarization. 

That said, there are strong tensions with the existing theoretical expectations in certain observables.
One is seen in figure~\ref{fig:PzSTAR}; the same hydrodynamic calculation that
  reproduced $\left\langle\vec{P}_{\Lambda}\cdot\hat{J}\right\rangle$ with no
  special tuning, predicts the wrong sign of the longitudinal polarization, $\left\langle\vec{P}_{\Lambda}\cdot\hat{z}\right\rangle$.
Hence it seems that, similar to the early collective flow studies, the framework is well-grounded, while
  there is much to learn from the details.
  Chapters 8~\cite{Karpenko:2021wdm}, 9~\cite{Huang:2020dtn} and 10~\cite{ch:Cao} in this Volume provide an in-depth discussion on the phenomenology study based on this framework. More broadly, the presence of global rotation has opened a new dimension for investigating its nontrivial effects, for example, on the phase structures of matter (see chapter 11~\cite{ch:Chen}) or on the interplay between orbital and spin angular momentum (c.f. chapter 12~\cite{Fukushima:2020qta}).

\subsection{Vector meson spin alignment - more complicated physics?}

\begin{figure}[t]
\centering
  \begin{minipage}[b]{0.45\textwidth}
    \includegraphics[width=\textwidth]{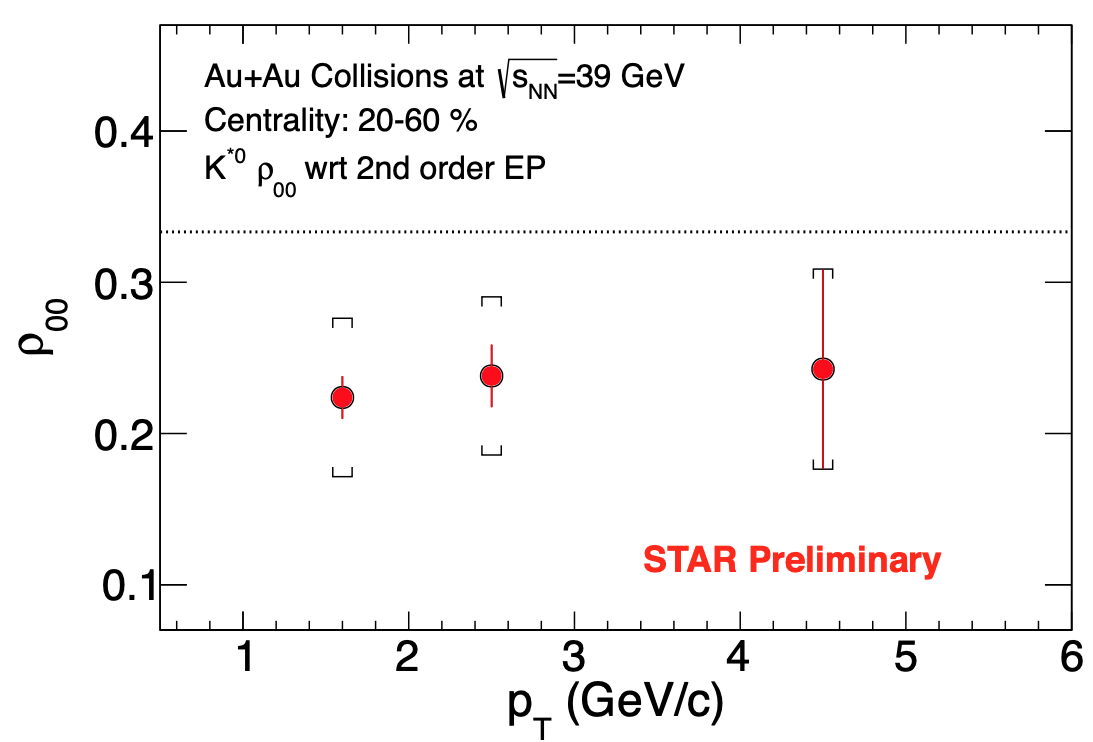}
  \end{minipage}
  \hfill
  \begin{minipage}[b]{0.45\textwidth}
    \includegraphics[width=\textwidth]{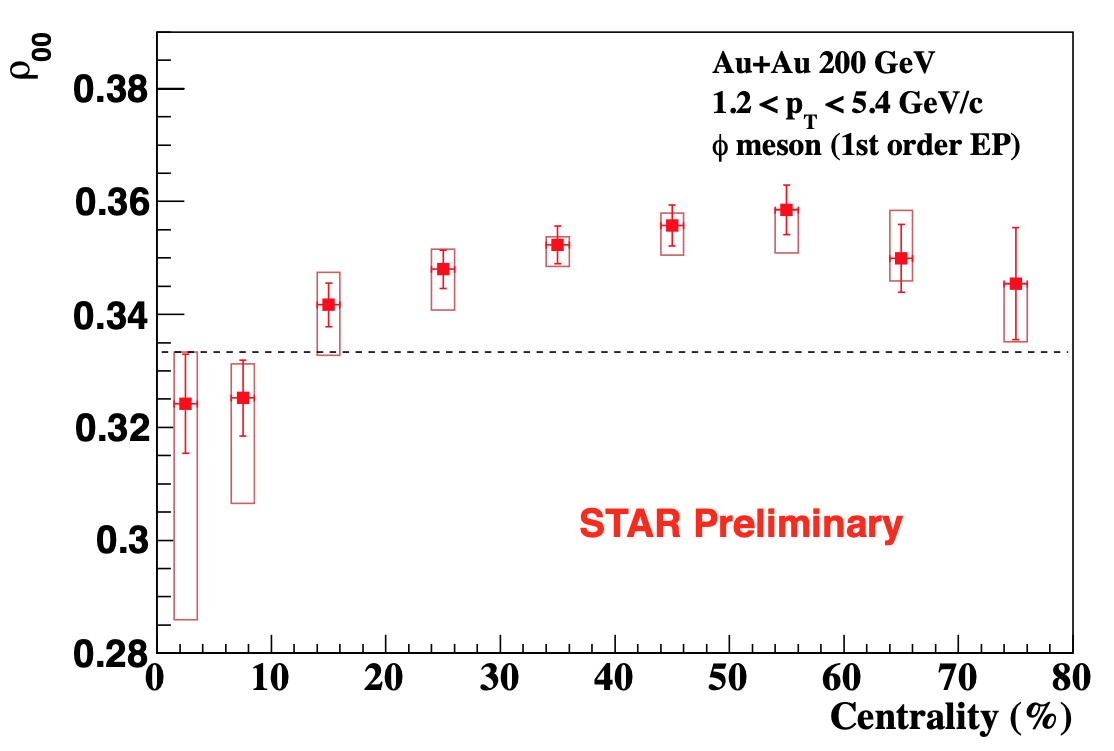}
  \end{minipage}
    \caption{\label{fig:STARmesons}
    Vector meson alignment in Au+Au collisions at $\sqrt{s_{NN}}=200$~GeV,
    measured by the STAR Collaboration at RHIC~\cite{Zhou:2019lun}.
    Left: $\rho_{0,0}$
    for $K^{*0}$ mesons as a function of transverse momentum
    for mid-central collisions. 
    Right: $\rho_{0,0}$
    for $\phi$ mesons as a function of centrality.
    Dashed lines indicate $\rho_{0,0}=\tfrac{1}{3}$,
    corresponding to no alignment with the normal to the event plane.}
\end{figure}

In an equilibrium picture, the spins of all emitted particles will be aligned
  with the total angular momentum of the system.
In addition to $\Lambda$ and $\overline{\Lambda}$ hyperons discussed above,
  preliminary results from the STAR collaboration indicate consistent polarization
  of $\Xi$, $\overline{\Xi}$ and $\Omega$ baryons~\cite{TakafumiPresentationAtRhicAUM2020}.
Besides baryons, in principle polarization could be detected for
  vector mesons such as ${\rm K}^{*}$ or $\phi$.

The spin of a vector meson is quantified by the $3\times3$ spin-density matrix
  $\rho_{i,j}$. Becattini discusses the coupling of this quantity to fluid vorticity in Chapter 2~\cite{Becattini:2020sww}. 
Due to the parity-conserving nature of their strong decay, the elements $\rho_{1,1}$
  and $\rho_{-1,-1}$ cannot be separately determined.
Because the trace is unity, there is only one independent diagonal element, $\rho_{0,0}$
  which quantifies the component of the meson spin perpendicular to the quantization axis.
As with the baryons, for the average spin, the axis of interest is $\hat{J}$,
  perpendicular to the event plane.
Random alignment of spins would yield $\rho_{-1,-1}=\rho_{0,0}=\rho{1,1}=\tfrac{1}{3}$.
Given only experimental access to $\rho_{0,0}$, it is impossible to determine whether
  the meson spin is parallel or anti-parallel to $\hat{J}$, but in either case,
  spin alignment would imply $\rho_{0,0}<\tfrac{1}{3}$~\cite{Leader:2001gr}.

The 2-particle decay topology of a vector meson is related to the alignment according 
to~\cite{Schilling:1969um}:
\begin{equation}
    \label{eq:SpinAlignment}
    \frac{{\rm d}N}{{\rm d}\cos\theta^*} = \frac{3}{4}\left[1-\rho_{00}+\left(3\rho_{00}-1\right)\cos^2\theta^*\right] ,
\end{equation}
where $\theta^*$ is the angle between the parent spin and a daughter momentum
in the parent's rest frame.
At local thermodynamic
equilibrium, the alignment is quadratic in thermal vorticity to first order~\cite{Becattini:2016gvu,Becattini:2020ngo}:
\begin{equation}
\label{eq:VectorMesonHydro}
    \tfrac{1}{3}-\rho_{00}\approx\tfrac{4}{9}\varpi^2 .
\end{equation}
Therefore, consistency with the hyperon results would lead to the expectation $\tfrac{1}{3}-\rho_{0,0}\approx10^{-3}$

\begin{figure}[t]
\centering
    \includegraphics[width=0.75\textwidth]{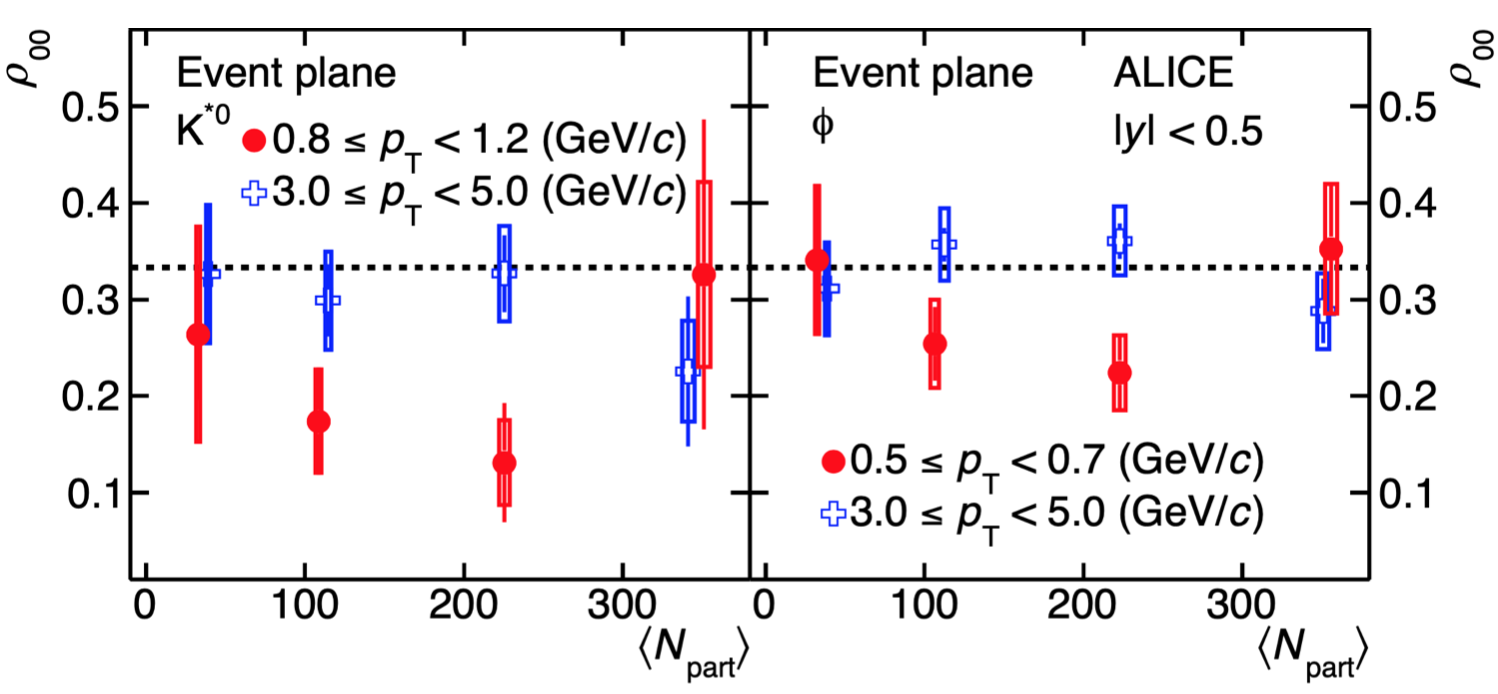}
    \caption{Vector meson alignment, for Pb+Pb collisions at $\sqrt{s_{NN}}=2.76$~TeV,
    measured by the ALICE Collaboration at the LHC~\cite{Acharya:2019vpe}.
    Left (right) panel shows $\rho_{0,0}$
    for $K^{*0}$ ($\phi$) mesons as a function of collision centrality, for two
    ranges in transverse momentum.  Dashed lines indicate $\rho_{0,0}=\tfrac{1}{3}$,
    corresponding to no alignment with the normal to the event plane.
    }
    \label{fig:ALICEmesons}
\end{figure}

Experimental results deviate strongly from that expectation.
Figures~\ref{fig:STARmesons} and~\ref{fig:ALICEmesons}  show $K^{0*}$ and $\phi$ alignment measurements from the STAR and
  ALICE experiments at RHIC and LHC, respectively.
In all cases, $|\tfrac{1}{3}-\rho_{00}|\approx0.1$, two orders of magnitude larger than expectations
  based on $P_{\Lambda}$ and vorticity considerations.
Perhaps more surprisingly, STAR reports $\rho_{00}>\tfrac{1}{3}$ for $\phi$ mesons.

As discussed above, the hydrodynamic equilibrium ansatz seems a reliable baseline for understanding
  hyperon polarization, as it is for understanding much else in heavy ion physics.
However, there may be many other effects at play.
In their original paper~\cite{Liang:2004xn}, Liang and Wang considered different hadronization
  mechanisms involving polarized quarks.
If vector mesons are produced by simple coalescence of a quark and antiquark with polarizations
  $P_q$ and $P_{\overline{q}}$, respectively, then
\begin{equation}
    \rho_{0,0}^{\rm meson}=\frac{1-P_{q}P_{\overline{q}}}{3+P_{q}P_{\overline{q}}} 
    \approx \tfrac{1}{3}-\tfrac{4}{9}\left(P_qP_{\overline{q}}\right)^2 ,
\end{equation}
where the approximation holds for small polarizations.
This is consistent with the hydrodynamic equilibrium prediction (equation~\ref{eq:VectorMesonHydro})
  if $P_{q}=P_{\overline{q}}=\varpi$.
It is not possible to reconcile the ALICE measurements of very small values of hyperon 
   with large values of $|\tfrac{1}{3}-\rho_{0,0}|$ in a simple recombination picture~\cite{Acharya:2019vpe}.

Perhaps even more surprising are measurements of the STAR Collaboration at RHIC.
For $K^*$, they report~\cite{Zhou:2019lun}  values of $\tfrac{1}{3}-\rho_{0,0}$
  similarly large as those seen at the LHC, but
  for $\phi$ mesons, $\rho_{0,0}>\tfrac{1}{3}$; c.f. figure~\ref{fig:STARmesons}.
Liang and Wang pointed out that hadronization via polarized quark fragmentation
  could result in $\rho_{0,0}>\tfrac{1}{3}$.
This mechanism may be most important at large rapidity or transverse momentum,
  but could in principle play a role at midrapidity, where these measurements
  are made.
However, naively, if fragmentation is the dominant hadronization mechanism, $K^*$ and $\phi$
  should be affected similarly.
Furthermore, it would seem natural that quark hadronization would be more important
  at LHC energies than at RHIC.

Sheng, Olivia and Wang~\cite{Sheng:2019kmk} propose that an entirely new physical
  effect could be at play, in which a hypothetical mean $\phi$ field couples to
  the system angular momentum.
Depending on the values of several parameters, $\rho_{0,0}^\phi$ could be greater
  or less than $\tfrac{1}{3}$.
In principle, by fine-tuning~\cite{Sheng:2019kmk} the energy dependence of four parameters, this
  model might accommodate $\rho_{0,0}^\phi>\tfrac{1}{3}$ at RHIC energies and 
  $\rho_{0,0}^\phi<\tfrac{1}{3}$ at the LHC.
Because this model is not expected to apply to $K^{*0}$ mesons~\cite{Sheng:2019kmk},
  it will be important to identify independent measurements that can constrain   and verify its assumptions.

In summary, it is clear that the situation with spin alignment of vector mesons is
  very different than that for hyperon polarization.
In the latter case, the equilibrium hydrodynamic paradigm which works well for
  other aspects of heavy ion collisions seems a reasonable starting point; polarization
  then allows a more sensitive probe of the system evolution at the finest scales.
For the vector mesons, however, it is clear that observations cannot be explained
  by this established paradigm.
Competing effects from multiple hadronization mechanisms, hadronic effects, and novel mean fields may
  be at play, differentially affecting the different particle species and different collision
  energies.
 See Chapter 7~\cite{Gao:2020lxh} by Gao, {\it et al} in this Volume, for an extensive discussion.
The phenomenon of vector meson spin alignment deserves continued intense theoretical focus;
  at the moment, the situation is too unclear to summarize what might be learned.

\subsection{Future experimental work}

Among the most pressing issues in the field of heavy ion physics is the existence and
  consequences of an intense, long-lived magnetic field.
Its presence could allow experimental access to novel effects due to chiral symmetry
  restoration.
Because $\Lambda$ and $\overline{\Lambda}$ have opposite magnetic moments, a strong
  $\vec{B}$-field at hadronization would lead to a polarization ``splitting''~\cite{Becattini:2016gvu,Muller:2018ibh,Guo:2019mgh}. 
The magnitude of the splitting remains below the statistical sensitivity of existing
  measurements, but the ongoing BES-II campaign at RHIC is expected to either discover
  the splitting or set meaningful limits on possible magnetic effects.
  
While the {\it average} (``global'') polarization must align with the total angular momentum
  of the collision, hydrodynamic and transport simulations predict a rich flow structure featuring
  nontrivial local vorticity.
The longitudinal polarization results in figure~\ref{fig:PzSTAR} represent the first observation
  of such an effect.
However, more complicated effects may be present on an event-by-event basis, leading to
  vorticity ``hot spots'' that may be revealed by spin-spin correlations~\cite{Pang:2016igs}.
Experimental searches for such a signal are ongoing at RHIC, but two-particle tracking
  artifacts make them highly challenging.
  
As discussed above, the physics driving vector meson spin alignment is apparently much
  more complicated than that behind $\Lambda$ polarization.
The STAR Collaboration at RHIC has presented a nearly finalized 
  study~\cite{TakafumiPresentationAtRhicAUM2020} of polarization of
  $\Xi$ and $\Omega$ hyperons which appear consistent with the $\Lambda$ polarizations,
  with small mass-dependent effects.

While the total system angular momentum decreases with reduced collision energy, the largest global
  polarization is observed at the lowest energy.
It will be important to measure polarization at still lower energies, below the energy
  threshold for QGP formation and at energy densities below the limits of applicability of
  hydrodynamics.
The BES-II program at RHIC includes a fixed-target campaign already producing
  results~\cite{Adam:2020pla} in this regime. 
Much higher statistics datasets at low energy are expected at the NICA and FAIR facilities
  soon to commence operation.

The energy dependence of the polarization signal may reflect an evolution of rotational flow
  structure away from midrapidity, where measurements have focused thus far, as the collision
  energy increases.
Experiments with good tracking near beam rapidity may probe strong vorticity resulting
  from the breakdown of longitudinal boost-invariance, challenging hydrodynamic and transport
  simulations more stringently than possible previously.

\section{Summary and Outlook}

Over several decades, the field of relativistic heavy ion physics has matured and focused on
  the creation and study of the quark-gluon plasma.
Hydrodynamics and transport theory have provided a useful paradigm in which to interpret
  a wide diversity of experimental results from high-energy collisions at RHIC and the LHC.
Theory and models based on this paradigm have become increasingly sophisticated,
  simulating the entire evolution of the dynamic system and making quantitative
  connection to the initial state and fundamental transport coefficients.

The observation of rotational phenomena has opened an exciting new direction into
  this well-developed and fertile environment, a rare example of a truly
  new development in a mature field.
First measurements of global hyperon polarization are largely consistent with predictions
  from existing hydrodynamic and transport simulations, indicating that the tools are
  at hand, to understand the phenomenon.
More differential measurements, of the azimuthal dependence of global and longitudinal
  hyperon polarization, are more difficult to understand; the effects have magnitudes
  in line with standard expectations, but reproducing the sign of the observed oscillations
  may require nontrivial revisions to our current understanding.
On the other hand, vector meson spin alignment-- presumably  related to hyperon polarization--
  is quantitatively and qualitatively impossible to understand solely in terms of the
  hydrodynamic paradigm that successfully explains other observables; here, there may
  be numerous competing effects that depend nontrivially on particle species and collision energy,
  including a newly-proposed coherent mesonic mean field.

Thus, it appears that the new phenomena of strongly interacting QCD matter under rotation
  may be addressed by current theory and models, while at the same time requiring new insights.
The contributions to this book represent a broad sample of some of the early theoretical
  efforts-- from fundamental theory to phenomenology-- to determine the physics behind these phenomena.
New insights are bound to result from continued theoretical focus and upcoming experimental
  results.
The following pages are the first chapters in what will surely be a much longer story.


\section*{Acknowledgements}
This work is supported in part by 
U.S. Department of Energy grant DE-SC0020651 and by U.S. National Science Foundation grant PHY-1913729.



\end{document}